\documentclass[prl,twocolumn,showpacs]{revtex4}
\usepackage{array,amsmath,amssymb,graphicx,cases}

\begin{document}
\title{\textsc{The Effect of Randomness on the Mott State}}

\author{Thomas Nattermann,  Aleksandra Petkovi\' c,
Zoran Ristivojevic, and Friedmar Sch\"utze
}

\affiliation{Institut f\"ur Theoretische Physik, Universit\"at zu
K\"oln, Z\"ulpicher Str. 77, 50937 K\"oln, Germany}

\date{\today}

\begin{abstract}

We reinvestigate the competition between
the Mott and the Anderson  insulator state
 in a one-dimensional disordered fermionic system by
a combination of instanton and renormalization group methods.
Tracing back both the compressibility and the ac-conductivity to a
vanishing kink energy of the electronic  displacement field we do
not find any indication for the existence of an intermediate (Mott
glass) phase.
\end{abstract}
\pacs{71.10Hf, 71.30.+h}

\maketitle

\emph{Introduction}. Two non-trivial routes to insulating behavior in solids
are connected with the names of Anderson and Mott. Non-interacting electrons in
disordered solids are  always localized in one and two space dimensions
resulting from  a combination of classical localization and quantum-mechanical
interference \cite{An_58,AbAnLiRa_79}. This Anderson insulator (AI) is
characterized by
 both a finite ac conductivity at low
frequencies and a finite compressibility \cite{MoHa_68}. Interaction between
electrons reduces the effect of disorder and may lead to metallic behavior in
two dimensions \cite{Fi_84}. In Mott insulators, on the other hand, insulating
behavior results from the blocking of sites by repulsive interaction between
electrons \cite{Mo_90} and hence are dominated by correlation effects.  The
Mott-insulating (MI) phase  is incompressible and has a finite gap in the
conductivity. A natural question is then: what happens in systems when both
disorder and interactions are non-negligible? Is there, depending on the
strength of interaction or disorder, a single transition between these two
phases or is the scenario more complex? A particularly interesting case to
answer these questions is that of electrons in one space dimension:
interactions are very strong, destroying Fermi liquid behavior, but
interactions can be partially incorporated into the harmonic (bosonized) theory
\cite{Gi_book}. Additional motivation to study this case comes from its
physical realization in quantum wires \cite{qw} and ultra-cold gases
\cite{ScMoScKoEs_04}.

One-dimensional systems of this type have indeed been considered in a number of
publications. Early work by Ma \cite{Ma_82} using real-space renormalization
group (RG) for the disordered 1D Hubbard model suggests a direct transition
from an AI to an MI phase. This result contrasts with more recent work (see
e.g. \cite{PaLiAn_93,OrGiDo_99} and references therein) where a new type of
order, different from the MI and AI phase, was found. In particular in
\cite{OrGiDo_99} the existence of a new \emph{Mott glass} (\emph{MG}) phase was
postulated which is supposed to be incompressible but has no gap in the optical
conductivity.

In this paper  we reinvestigate this
problem by relating  both the
compressibility and the low frequency
conductivity to the energy of kinks in the
displacement pattern of the bosonized
electrons, resorting to ideas known from
the discussion of the roughening
transition \cite{No_92}.  Using an
instanton calculation of the conductivity
\cite{NaGiDo_03,RoNa_06} we show that an
incompressible phase has always a gap in
the low frequency conductivity, excluding
the possibility of a MG phase. A number of
further consideration support this
finding.

\emph{The model}. Following Haldane
\cite{Ha_81} we relate the mass  density
$\rho(x)$ of the electrons to their
displacement field $\varphi(x)$:
$\rho(x)=\rho_0+\frac{1}{\pi}\partial_x\varphi+
\rho_0\cos(2\varphi+2k_Fx)+\ldots$.
Throughout the paper we will use  $e=\hbar=1$. The ground state
energy of the system follows from $ E_0=-\lim_{T\to 0}T\ln\int
D\varphi e^{-S}$ where the action is \cite{Gi_book, OrGiDo_99}
\begin{eqnarray}
  &{S}&= \frac{1}{2\pi K }
    \int\limits_0^{L\Lambda} \int\limits_{0}^{\lambda_T\Lambda}dx
    d\tau
    \Big\{(\partial_{{\tau}}\varphi)^{2}+
    (\partial_{x}\varphi + \mu(x))^{2}- \label{eq:full_action} \\
   &&-\frac{2\pi\kappa}{\Lambda^{2}} F\varphi-(\zeta(x)e^{-2\varphi
   i}+h.c.)
     -w\cos
   2\varphi\Big\}+S_{\textrm{diss}}.\,\,\,\,\nonumber
\end{eqnarray}
Here we  introduced dimensionless space
and imaginary time coordinates by the
transformation $\Lambda x\to x$ and
$\Lambda v \tau \to \tau$ where $v$ is the
plasmon velocity and $\Lambda$ a large
momentum cut-off.  $\lambda_T=v/T$ and
 $L$ and denote the thermal de Broglie wavelength
of the plasmons and the system size,
respectively. $w$ denotes the strength of
the Umklapp scattering term (or the
strength of a commensurate potential) and
$F$ the external driving force. $\mu(x)$
and $\zeta(x)$ result from the coupling of
the random impurity potential to the long
wave length and the periodic part of the
density, respectively, with
 $   \langle
    \zeta(x)\zeta^*(x')\rangle=u^2\delta(x-x'),
    \,\, \langle\mu(x)\mu(x')\rangle =
    \sigma\delta(x-x')$,
all other correlators  vanish. 
Finally we add a
dissipative action $S_{\textrm{diss}}$
describing an Ohmic resistance  which may
e.g. result from the interaction with an
external gate \cite{Gu_06}.

\emph{Rigidities.} We begin with a discussion of the generalized
rigidities, which are related
 to the compressibility and the
conductivity of the system. We first consider the application of a
fixed strain $\vartheta$ by imposing the boundary conditions
$\varphi(0,\tau)=0$ and $\varphi(L,\tau)=\pi\vartheta L \Lambda$.
The boundary condition in the $\tau$-direction is assumed to be
periodic. For $ \vartheta\ll 1$ and $L\to \infty $ the corresponding
increase $\Delta E_0(\vartheta,0)= E_0(\vartheta,0)- E_0(0,0)$ of
the ground state energy $E_0(\vartheta,0) $ is clearly an even but
not necessarily analytic function of $\vartheta$. Thus
\begin{numcases}{\!\!
\frac{\Delta
E_0(\vartheta,0)}{L}\Big|_{L\to\infty
}\approx}
 \Upsilon_x{\vartheta}^2/2\,\,\,\,\,\,& \textrm{if}
 $\Sigma_x=0,$
\label{eq:Upsilon_x}\\
\Sigma_x|\vartheta|&\textrm{if}
$\Upsilon_x^{-1}=0.$ \label{eq:Xi_x}
\end{numcases}
The r.h.s. of this relation has to be understood  as follows: if
$\Sigma_x=0$ then the stiffness $\Upsilon_x=\Lambda^2/\kappa$
describes the response to the twisted boundary conditions,
$\Upsilon_x^{-1}$ is the isothermal compressibility
\cite{NaGiDo_03}. In this case the change of $\varphi$ is spread
over the whole sample. If however $\Upsilon_x $ diverges, i.e. the
system becomes incompressible, then the kink energy
$\Sigma_x/\Lambda$
 is non-zero.
In this case the change of $\varphi$ from $0$ to $\pi$ occurs in a
narrow kink region of width $\xi$ much smaller than $L$. Creating a
kink corresponds to adding (or removing) an electron at the kink
position.  A non-zero kink energy resembles the step free energy
 of a surface below the roughening
transition  \cite{No_92} \footnote{If we apply instead of the fixed
boundary conditions an external stress to the system, then
$\Sigma_x$ is of the order of the critical stress to generate the
first kink.}.

In a similar manner we can  apply non-trivial boundary conditions in
the $\tau$-direction by choosing $\varphi(x,0)=0$ and
$\varphi(x,L_{\tau})=\pi j L_{\tau}/v$. This corresponds to imposing
an external current  $j=\langle
\partial_{\tau} \varphi\rangle/\pi$ at $x=0$ and $x=L$:
\begin{numcases}{\frac{\Delta E_0(0,j)
}{L}\Big|_{L\to\infty}\approx}
 \Upsilon_{\tau}j^2/2\,\,\,\,& if $\Sigma_{\tau}=0$,
\label{eq:Upsilon_y}\\
\Sigma_{\tau}|j| & if
$\Upsilon_{\tau}^{-1}=0$\label{eq:Xi_y}.
\end{numcases}
$\Upsilon_{\tau}$ is related (at $T=0$) to
the charge stiffness $D=1/\Upsilon_{\tau}
=\kappa v^2 $. $D$ determines the Drude
peak of the conductivity $    \sigma
    (\omega)=D\delta(\omega)+
    \sigma_{\textrm{reg}}(\omega)$ \cite{Gi_book,Ko_64}.
In Lorentz invariant systems (like the MI)
$\Upsilon_x=(\Lambda v)^2\Upsilon_{\tau}$
and $\Sigma_x=\Lambda v\Sigma_{\tau}$,
provided they are finite.
\begin{figure}
\centerline{
\includegraphics[width=0.8\linewidth]{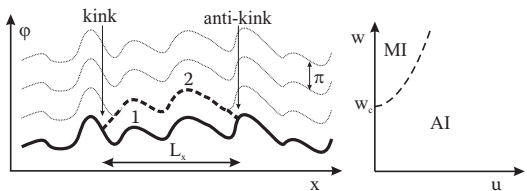}
}{\caption{Left: Kink and anti-kink in the displacement profile
$\varphi(x)$. The thin lines represent the minima of the potential
energy in the absence of the driving force, the bold line one
metastable state and the dashed line  the instanton configuration,
respectively.  Applying a fixed strain to the system only kinks (or
anti-kinks) are enforced in the system. Right: schematic phase
diagram with $w_c\sim\sigma^2$. }\label{Fig:wall} }
\end{figure}

So far we assumed that
$\Upsilon_{x/\tau},\,\Sigma_{x/\tau}$ are
self-averaging if $L\to \infty$. In the
same way we may introduce \emph{local}
rigidities  by applying twisted boundary
conditions over a large but finite
interval $[x,x+L_x\Lambda]$,
$\Lambda^{-1}\ll L_x\ll L$. The result for
$\Upsilon_{x/\tau}$ and $\Sigma_{x/\tau}$
will then depend  in general on the size
$L_x$ of the 
interval , i.e.  $\Sigma_{x/\tau}\to
\Sigma_{x/\tau}(L_x)$ .

\emph{Conductivity}. In cases where $D$
vanishes, the  field or frequency
dependent conductivity
 $\sigma(F,\omega)$ may still be non-zero,
 provided $F$ or $\omega$ are finite.
The \emph{non-linear dc-conductivity} $\sigma(F,0)$
 follows from an instanton approach in which the
current is calculated
 from the tunneling probability of electrons between
consecutive metastable states (see
Fig.\ref{Fig:wall}).  The latter are the
classical ground states in the absence of
the driving force which follow from each
other by a shift of a multiple of $\pi$.
This approximate treatment is reliable for
not too large $K$ and small $F$. The
conductivity $ \ln
 \sigma(F,0)=-S_{\textrm{I}}( L_{x,s},
 L_{\tau,s})
$ is then related to the action $S_{\textrm{I}}$ of a critical droplet (the
instanton) of the new metastable state on the background of the old one
\cite{NaGiDo_03,RoNa_06}. Here $ L_{x,s}, L_{\tau,s}$ are
 the saddle point values of the extension of the instanton
  which follow from
 \begin{equation}\label{eq:instanton}
    S_{\textrm{I}}\approx
    \,\frac{2\Sigma_x(L_x)}{\Lambda v}L_{\tau}+
    L_x(2\,\Sigma_{\tau}+\eta\ln\frac{L_{\tau}}{\xi})-
    \frac{1}{\pi v}F L_xL_{\tau}.
 \end{equation}
For simplicity we assumed that the instanton is rectangular, a
choice which is motivated by the fact that the disorder is
correlated in time \cite{NaGiDo_03}.   $\Sigma_x/(\Lambda v)$ and
$\,\Sigma_{\tau}$ play the role of  the surface tension of the
instanton. The last term is the bulk contribution from the external
field. The $\eta$-term results from the dissipation which allows
relaxation in the next metastable state \cite{NaGiDo_03}.

The low frequency \emph{ac-conductivity} $\sigma(0,\omega)$ results
from the spontaneous tunneling processes between metastable states
and their instanton configurations (1 and 2 in Fig. \ref{Fig:wall}).
Tunneling leads to a level splitting of the order
\begin{equation}\label{eq:ac_conductivity}
    \delta E\approx \sqrt{{4\Sigma_x^2(L_x)}{\Lambda^{-2}}
+{C}{(v\Lambda \xi K)^2}e^{-4L_x\Sigma_{\tau}}},
\end{equation}
which has to match the energy $\omega$ of the external field
\cite{RoNa_06}. $1/\Sigma_{\tau}\sim K\xi$ plays the role of the
tunneling length, $C$ is a numerical factor. This mechanism was
first considered for non-interacting electrons by Mott
\cite{MoHa_68} and extended to the interacting case via instantons
in  \cite{RoNa_06}. Thus, to get a non-zero $\sigma(0,\omega)$ for
{arbitrary} low frequency $\omega$, $2\Sigma_x(L_x)<
\omega\Lambda\to 0$ which requires  $2\Sigma_x(L_x)\to 0$ for a
finite density of kink positions, which implies a finite
compressibility. The distance $L_x$ between the sites involved in
the tunneling is then $L_x({\omega})\approx \xi
K\ln(1/(\kappa\xi\omega))$.

{\emph{Fixed points and phases.}} We come
now to the discussion of the possible phases of model
(\ref{eq:full_action}) by attributing them to their  RG fixed points
(denoted by superscript $^*$). Bare values will get a subscript
$_0$. For small $u$ and $w$ the lowest order RG-equations read
\cite{He_98}:
\begin{eqnarray}
\frac{d K}{dl}&=&
-K(au^2+bw^2),\,\,\,\,\,\frac{d
\sigma}{dl}=\sigma(1-cw^2)\label{eq:RG-flow3}\\
\frac{d w}{dl}&=& w(2-K-\frac{2}{\pi}\sigma) \label{eq:RG-flow1}\\
\frac{d u^2}{dl}&=& u^2(3-2K)+\frac{1}{\pi}\sigma w^2 \label{eq:RG-flow2}\\
\frac{d\kappa}{dl}&=&-\kappa(z-1+\frac{c}{2}w^2) \label{eq:RG-flow4}
\end{eqnarray}
where $l$ is the logarithm of the length scale and $a$, $b$, $c$
are positive non-universal constants. $z$ denotes the dynamical
critical exponent which has to be determined from the fixed point
condition.

\emph{Luttinger liquid (LL).} The LL phase
is characterized by  $u_L^*=w_L^*=0$ and
hence  $\Sigma_x=\Sigma_{\tau}=0$. $z=1$,
$K^*_L>0$ and $\kappa_L^*=K_L^*/(\pi
v_L^*)>0$. The fixed point is
 reached for sufficiently large values of
$K$ and $\sigma$. The long time and large scale behavior of the
system is that of a clean LL characterized both by a finite
compressibility $\kappa_L^*$ and a finite charge stiffness
$D_L=\kappa_L^*v_L^{*2}$. The dynamical conductivity is given by
$\sigma_{\emph{reg}}=iD_L/({\pi \omega})$.
 The presence of the forward scattering
term $\sim \mu(x)\partial_x \varphi$ does not change these results
since it can be always removed by the transformation
$\varphi=\tilde\varphi-\int_0^xdx'\mu(x')$.


\emph{Mott insulator (MI)}. Here $K^*_M=\kappa^*_M=u^*_M=\sigma^*_M=0$ but
$w^*_M\gg 1$. Clearly the fixed point $w_M^*$ is outside the applicability
range of the (\ref{eq:RG-flow3})-(\ref{eq:RG-flow4}) but nevertheless some
general properties of this phase can be concluded. The system is in the
universality class of the $2$-dimensional classical sine Gordon model which
describes inter alia the MI to LL transition and the roughening transition of a
2-dimensional classical crystalline surface \cite{No_92}. In the MI phase
$\Upsilon_x$ and $\Upsilon_y$ diverge. The classical ground state is given by
$\varphi(x)= n\pi$ with $n$ integer. The system is characterized by a finite
kink energy $\Sigma_x \sim \Lambda(\kappa_0\xi_M)^{-1}$ where $\xi_M$ denotes
the correlation length of the MI-phase
\cite {No_92}. 
From (\ref{eq:instanton}) we get $\sigma(F,0)\sim \exp(-F_M/(KF))$
 where $F_M\approx 4\pi \Sigma_x\Sigma_{\tau}/\Lambda \sim 1/(\kappa_0\xi_M^2)$ is a
characteristic depinning field \cite{Ma_77}. According to
(\ref{eq:ac_conductivity}) the ac conductivity vanishes for
$\omega\lesssim 2\Sigma_x/\Lambda$.

\emph{Anderson insulator (AI)}. Here $w_A^*=K_A^*=0$ but $u_A^*\gg 1$.
$\kappa_L^*\approx\kappa_0$ is finite which is the result of the so-called
statistical tilt symmetry \cite{Schultz_88} corresponding to $z=1$. The fixed
point Hamiltonian is in the universality class of the $1+1$ dimensional sine
Gordon model with a random phase correlated in the $\tau$-direction.  The
transition to the LL phase occurs at $K=K_{A}^*(u)\ge 3/2$.

Next we look at $\Sigma_x(L_x)$ \emph{finite} length scale $\xi_A\ll
L_x\ll L$  such that the parameters are close to their fixed point
values.  $\xi_A$ is the correlation length which diverges at
the Beresinskii-Kosterlitz-Thouless (BKT) transition to the LL phase. 
To find the  classical ground state of the system under periodic boundary
conditions we have first to choose $2\varphi_i+\alpha_i=2\pi n_i$ with $n_i$
integer, and secondly the elastic term has to be minimized with respect to the
$n_i$. The subscript $i$ refers to the sites of the lattice with spacing
$\xi_A$ and $\zeta_i=|\zeta_i|e^{i\alpha_i}$. The solution is $n_i=\sum_{j\le
i}[({\alpha_j-\alpha_{j-1}})/{2\pi}]_G$ \cite{NaGiDo_03}. $[x]_G$ denotes the
closest integer to $x$. The ground state is uniquely determined by the
$\alpha_j$ apart from the pairs of sites (of measure zero) at which
$\alpha_j-\alpha_{j-1}=\pm \pi$. At such pairs the ground state
\emph{bifurcates} since two solutions are possible. In the case of
non-interacting electrons ($K=1$) bifurcation sites are states \emph{at} the
Fermi energy. For pairs at which $\alpha_j-\alpha_{j-1}=\pm
\pi+\epsilon,\,|\epsilon|\ll 1$ we can go over to an excited state by creating
a kink which costs at most the energy $\Sigma_x\approx\epsilon \Lambda/ (\kappa
\xi_A)$ \cite{NaGiDo_03}. Those 'almost' bifurcating sites correspond to states
close to the Fermi energy. The smallest $\epsilon$ found with probability of
order one in a sample of length $L_x$ is of the order $\xi/L_x$ and hence
$\Sigma_x\approx\Lambda /( L_x\kappa_0)$, i.e. the kink energy vanishes for
$L_x\to \infty$ \cite{NaGiDo_03} and hence the system is compressible.

Twisted boundary conditions in the $y$-direction give  $\Sigma_y
\sim (K_0\xi_A)^{-1}$ \cite{NaGiDo_03}. The non-linear conductivity
is given by $ \sigma(F, \omega=0)\sim
    \exp{-\sqrt{F_A/(FK^2)}}$
 where $F_A\sim{1/(\kappa
\xi_A^2)}$ \cite{NaGiDo_03}. This result can be understood as
electron tunneling between sites at which $\Sigma_x\sim 1/L_x$.  As
explained already a vanishing $\Sigma_x$ is also crucial for the
existence of the low frequency conductivity
$\sigma(0,\omega)\approx\xi_MK \left( \omega
    \kappa_0K\xi_M\ln({{\kappa_0\xi_M\omega}})\right)^2$
as has been discussed in detail in \cite{ RoNa_06}. This result can
be understood in terms of tunneling processes between \emph{rare}
positions at which the kink energies $\Sigma_x(x)$ are much smaller
than $1/(\kappa_0 L_x)$.
\begin{table}[htb]
\caption{\label{tab} Properties of
phases.}
\begin{ruledtabular}
\begin{tabular}{c c c c c c c}
Phase & $\kappa^*$&$K^*$&$D$&$\Sigma_x$&
$\Sigma_y$&$\sigma(\omega \ll \Sigma_x)$  \\
\hline
LL &$\kappa_L>0$&$K_L>0$&$\kappa_Lv_L^2$&$0$&$0$&$iD_L/(\pi\omega)$\\
 \emph{AI} & $\kappa_A>0$&$$0$$&$0$& $0$ &$ \sim\xi_A^{-1}$ & $\sim \omega^2\ln^2\omega$\\
\emph{MI} &$0$&$0$&$0$ &$ \sim\xi_M^{-1}$&$ \sim\xi_M^{-1}$ & $0$ \\
\end{tabular}
\end{ruledtabular}
\end{table}

\emph{Mott-glass (MG)}. This new hypothetical phase was proposed in
\cite{OrGiDo_99} to be characterized by a vanishing compressibility,
$\kappa^*_G=0$, but a non-zero optical conductivity at low frequencies
\cite{OrGiDo_99}. Since the phase is considered to be glassy, both fixed point
values $w_G^*, u_G^*\gg 1$. Similarly to the AI, the ground state can be found
by minimizing first the two backward scattering terms followed by minimization
of  the elastic energy. Although the ground state solution is now more involved
than for the MI and AI case, for $F=0$ it is clearly periodic with period
$\pi$. As before kinks (or anti-kinks) with $\delta\varphi=\pm \pi$ allow the
accommodation of twisted boundary conditions and the formation of instantons. A
vanishing compressibility corresponds to a \emph{finite} kink energy
$\Sigma_x\ge{\cal C}>0$ which, according to (\ref{eq:ac_conductivity}), leads
to a gap in the ac-conductivity. Here we make the reasonable assumption that it
is the instanton mechanism which dominates the low frequency response
\cite{Fo_02}. Thus, in a system with a non-zero $\sigma(\omega)$ for small
$\omega$ also the compressibility has to be non-zero, contrary to the claims in
\cite{OrGiDo_99}.

\emph{Phase diagram}.  We come now to the
discussion of the phase diagram of our
model (\ref{eq:full_action}).
 From
(\ref{eq:RG-flow2}) follows that the random backward scattering term is
generated by forward scattering and the commensurate pinning potential. Since
$\sigma(l)=\sigma_0e^l$ the two Eigenvalues $\lambda_1= 3-2K$ and
$\lambda_2=4-2K-4\sigma/\pi$ describing the RG-flow of $u^2$ and $w^2$ around
the the LL fixed point $u^*=w^*=0$ have opposite sign: $u(l)$ increases whereas
$w(l)$ decreases. Thus the hypothetical MG phase, if it existed, cannot reach
up to the point $u=w=0$, in contrast to the findings in \cite{OrGiDo_99}. From
this we conclude that for not too large values of $w_0$ the AI phase is stable,
as shown in Fig.~\ref{Fig:wall}. To find the phase boundary to the MI phase we
consider the stability of the MI phase with respect to the formation of a kink
by the disorder. To lowest order in the disorder we get for the kink energy in
the MI phase
\begin{equation}\label{eq:Sigma}
     \Sigma_x \sim \frac{\sqrt
    w_0}{\kappa_0}
    \big[1-\frac{1}{2}({\sigma_0^2}/{w_0})^{\frac{1}{4}}-
    {u_0}/{(\pi^2w_0^{3/4})}\big]
\end{equation}
 which gives for the phase boundary between the MI
and the AI phase as depicted in  Fig.~\ref{Fig:wall}. A similar result follows
from the self-consistent harmonic approximation.

So far we considered only typical disorder fluctuations. If
$|\zeta(x)|$ and $\mu(x)$ are Gaussian distributed and rare events
are taken into account, then (\ref{eq:Sigma}) remains valid with the
replacements $\sigma_0\to \sigma_0\ln(L\Lambda)$ and $u^2_0\to
u_0^2\ln(L\Lambda)$. Thus the size of the \emph{MI} phase is reduced
but finite unless $L\to\infty$.

 These findings are supported by the
observation that the forward scattering term in (\ref{eq:full_action}) can be
removed from the action by the transformation
$\varphi(x)=\tilde\varphi(x)-\int_0^{x}dx'\mu(x')$ \cite{FiWeGrFi_89}. On
sufficiently large scales $\gtrsim 1/(\sigma_0\Lambda)$ this transformation
leaves \emph{only  } random backward scattering term of strength
$\sqrt{u_0^2+w_0^2/(4\sigma_0)}$ in the action, which results for sufficiently
small $K$ in the AI-phase. Thus, to conlude we only find the three phases LL,
MI and AI.

Finally a word on the variational approach and replica symmetry breaking (RSB)
which has been used in \cite{OrGiDo_99} where the Mott glass  phase has been
found. In the variational approach the full Hamiltonian is replaced by a
harmonic one which leads to the decoupling of Fourier components $\varphi_q$
with different wave-vector $q$. Without RSB one obtains from this approach  the
results of perturbation theory which is valid only on small scales. RSB gives
the possibility of a further reduction of the free energy. The results obtained
in this way are exact only in cases when the coupling between different Fourier
components is irrelevant and the physics is dominated by the largest length
scale. Thus RSB is not an intrinsic property of the true solution of the
problem, but a property of the variational approach. A good example is the
related problem of the interface roughening transition in a random potential
\cite{EmNa_98}. The variational approach with RSB gives three phases: a flat, a
rough  and a glassy flat phase \cite{BoGe_92}, where the rough phase has Flory
like exponents. The functional RG (which takes the coupling of different
Fourier modes into account) gives however only two phases: the flat and the
rough phase, but \emph{at} the transition a logarithmically diverging interface
width \cite{EmNa_98}. A similar situation may exist also in the present case
although we are not able to calculate the properties of the AI to MI
transition.

To conclude we have shown for a one-dimensional disordered Mott
insulator (\ref{eq:full_action}), tracing back both the
compressibility and the ac-conductivity to the kink energy
$\Sigma_x$ of the electronic displacement field  that an
incompressible system has also a vanishing optical conductivity.
Thus we exclude the possibility
 of the existence of a Mott-glass phase.

We acknowledge helpful remarks by Sergey Artemenko and Achim Rosch
and the SFB 608 for financial support.


\end{document}